\input{aipcheck}

\documentclass[
    ,final            
  ]
  {aipproc}

\layoutstyle{6x9}

\newcommand{\beq}{\begin{equation}}
\newcommand{\eeq}{\end{equation}}
\newcommand{\beqa}{\begin{eqnarray}}
\newcommand{\eeqa}{\end{eqnarray}}
\newcommand{\Eq}[1]{(\ref{#1})}
\newcommand{\nm}{\nonumber\\}

\newcommand{\la}{\langle}
\newcommand{\ra}{\rangle}

\begin{document}

\title{Random and deterministic walks on lattices}

\author{Jean Pierre Boon}{
  address={Physics Department, Universit\'{e} Libre de Bruxelles,
                  1050 - Bruxelles, Belgium\\
                  E-mail: {\tt jpboon@ulb.ac.be}; URL: {\tt www.poseidon.ulb.ac.be}}
}



\begin{abstract}

        Random walks of particles on a lattice are a classical
        paradigm for the microscopic mechanism underlying
        diffusive processes. In deterministic walks, the role of
        space and time can be reversed, and the microscopic
        dynamics can produce quite different types of behavior
        such as  directed propagation and organization, which
        appears to be generic behaviors encountered in an
        important class of systems. The various aspects of 
        classical and not so classical walks on latices are
        reviewed with emphasis on the physical phenomena that
        can be treated through a lattice dynamics approach.




\end{abstract}

\maketitle


\section{Temporal versus spatial dispersion}

One of the fundamental physical paradigms, applicable to a wide variety of
physical processes, is that of {\em spatial diffusion}. The text-book example  
is a random walker on a one-dimensional lattice (see, e.g. \cite{feller}) where 
at each tick of the clock, the walker takes a step either to the left or to the right, 
the direction being chosen randomly with equal probabilities. One then asks 
what is the probability that the walker be at a given position after a given time.
If the walker starts at a known point, the answer is a binomial distribution
which, in the continuum limit, becomes a Gaussian. The variance of the
Gaussian grows with time so that the localization of the walker decreases,
and we say that the walker disperses. If the probability for the walker to
step in one direction is greater than that for the opposite direction, then
the walker propagates in the direction of higher probability and will
eventually visit each site of the lattice in that direction. The typical spatial
diffusive behavior is then manifested in the continuum limit as a Gaussian 
about a most-likely position which moves at a constant velocity. However, 
there are a number of situations in which, instead of asking where the
walker would be after a given time (long with respect to the duration of
an elementary time step), it is more natural to ask how long it will take to 
reach a given point, at some large distance from the starting position
(large compared to the unit length covered during the elementary time step).
More precisely for a stochastic process, one then asks what is the 
distribution of times taken to reach that point, a question related
to the problem of {\em first-passage processes} \cite{redner}. 

Everyday examples involve processes in which the goal is to arrive at a 
given point: for example, the marathon (wherein we ask for the distribution 
of finishing times), certain financial instruments, such as stock options 
(wherein we ask for the distributions of times needed for an asset to reach 
a certain value), traffic-flow (wherein we ask for the distribution of arrival times 
at destination), and packet transport over the internet. 
A more technical example is the behavior of certain cellular automata which 
model the motion of a particle on a substrate of scatterers (in 1 or 2 dimensions) 
where, for certain types of scatterers, the particle ends up propagating along a
particular channel, and, again, the first-passage time is the physical quantity 
of interest. A paradigm for this type of behavior is the automaton known as 
"Langton's ant" \cite{ant, cohen} which is described below. The interesting
fact is that even a simplified one-dimensional version of that automaton shows
the same type of behavior. This 1-D model is the analogue of the
one-dimensional random walker, with the important difference that the roles of 
space and time are reversed: for large distances, the distribution of first-passage 
times is Gaussian in the time variable with a variance that grows with increasing 
distance from the origin. In analogy with spatial diffusion that occurs in ordinary 
diffusive phenomena, this generic behavior is called \emph{temporal diffusion}.
We will show how starting form a simple model, a general 
{\em first-visit equation} is obtained which in the hydrodynamic limit yields
the {\em propagation-dispersion equation} (PDE), the analogue of the classical
advection-diffusion equation, and how this PDE further generalizes 
propagation and dispersion processes. 

\section{First visit equation}

The automaton known as "Langton's ant" \cite{ant}  lives in a two-dimensional
universe spanned by the square lattice with checker board parity,
so defining H sites and V sites. A particle (the ant) moves from site to site 
(by one lattice unit length) in the direction given by an indicator. One may
think of the indicator as a `spin' (up or down) defining the state of the 
site. When the particle arrives at a site with spin up (down), it is 
scattered to the right (left) making an angle of $+\pi/2$ ($-\pi/2$) with 
respect to its incoming velocity vector. But the particle modifies the
state of the visited site (spin up $\Longleftrightarrow$ spin down) so that 
on its next visit, the particle is deflected in the direction opposite to the 
scattering direction of its former visit. Thus the particle entering from 
below a H site with spin up is scattered East, and on its next visit to 
that same site (now with spin down), if it arrives from above, it will be 
scattered East again, while if it arrives from below, it will be scattered
West. Similar reasoning shows how the particle is scattered North or South
on V sites.

At the initial time, all sites are in the same state (all spins up or down),
and the position and velocity direction of the particle are fixed, but arbitrary.
So if we paint the sites black or white according to their spin state and we
start with say an all white universe, then, as the particle moves, the visited 
sites turn alternately black and white depending on whether they are visited 
an odd or even number of times. This color coding offers a way to observe 
the evolution of the automaton universe.  
The particle starts exploring the universe by first creating centrally 
symmetric transient patterns (see figures in references \cite{ant}), 
then after about 10 000 time steps (9977 to be precise), 
it leaves a seemingly `random territory'  to enter a `highway' 
(see Fig.1) showing a periodic pattern. The "disordered"
phase is not what a random walk would produce: the automaton is deterministic 
and its rules create correlations between successive states of the substrate, 
so also between successive positions of the particle. The power spectrum 
computed from the particle position time correlation function measured over 
the first 9977 time steps goes like $\sim \nu^{-\zeta}$ with 
$\zeta \simeq 4/3$. In the ordered phase (the `highway'), the power spectrum 
shows a peak at $\nu = 1/104$ with harmonics. Indeed in the highway, the particle 
travels with constant propagation speed: $c = 2 \sqrt 2/104$ (in lattice units).

\begin{figure}
 \includegraphics[height=.7\textheight]{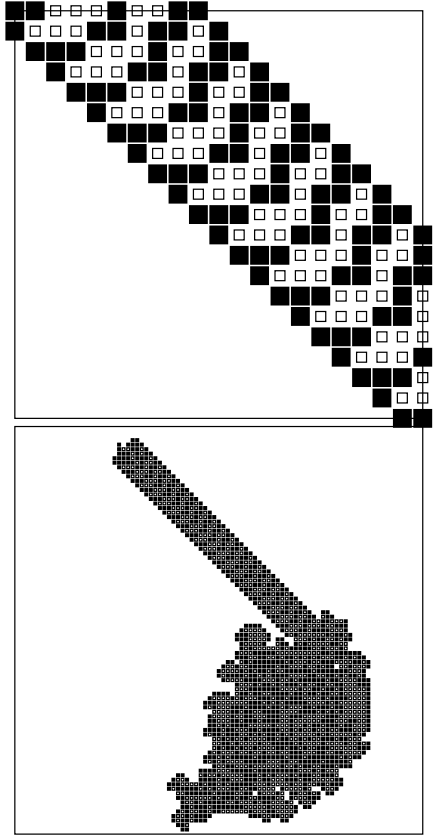}
 \caption{
 Langton's ant trajectory after 12,000 automaton time steps. The upper
 box is a blow-up of the highway showing the periodic pattern. Sites with
 open squares and dark squares have opposite spin states (up and down). }
\end{figure}

Because of the complexity of the dynamics on the square lattice, Grosfils,
Boon, Cohen, and Bunimovich \cite{grosfils} developed a one-dimensional
version of the automaton for which they provided a complete mathematical
analysis also applicable to the two-dimensional triangular lattice. In the 
one-dimensional case, the particle moves in the direction of its velocity vector
with probability $q$ and in the opposite direction with probability $(1-q)$, the 
direction being dictated by the "spin" of the lattice site, which is then reversed 
after the passage of the particle. The mean-field equation describing the
microscopic dynamics of the particle with the general condition
that the spins at the initial time are randomly distributed on the lattice, reads
\cite{grosfils}
\beq
\label{1D_eq}
f(r+1,t+1) \,=\, q\, f(r,t)\,+\,(1-q)\, f(r,t-2)\,.
\eeq
Here $f(r,t)$ is the single particle distribution function, i.e. the
probability that the particle visits site $r$ {\em for the first time}
at time $t$, and $q$ is the probability that the immediately previously
visited site along the propagation strip (the highway) has initially 
spin up, i.e. the probability that the particle be scattered along the direction 
of its velocity vector when arriving at the scattering site at $r-1$. 
\footnote{A similar equation holds for the two-dimensional triangular lattice 
and these equations were shown to yield exact solutions for
propagative behavior (corresponding to an ordered phase of the lattice) in 
the classes of models considered by Grosfils {\em et al.} \cite{grosfils}.}
An important result follows that can be formulated as a theorem \cite{grosfils} :
a particle moving from site to site in a one-dimensional lattice 
fully occupied with flipping scatterers (spins), propagates in one 
direction, independently of the initial distribution of the spins on the 
lattice.

There are two points of particular interest here. \\
(i) First one notices that 
Eq. (\ref{1D_eq}) has the same structure as the equation
for the one-dimensional random walk \cite{feller} 
\beq
\label{RW_eq}
g(r+1,t+1) \,=\, q\, g(r,t)\,+\,(1-q)\, g(r+2,t)\,,
\eeq
except that the in the second term on the r.h.s. of Eq. (\ref{1D_eq})
one has $t-2$ whereas in the random walk equation  (\ref{RW_eq})
one has $r+2$; this increment transfer between space and time makes 
a crucial difference as we shall see below.\\
(ii) Eq.(\ref{1D_eq}) is a particular case of a general equation \cite{boon}. 
To see this, consider a walker on a one-dimensional lattice and let ${f}(t/\delta
t;r/\delta r)$ be the probability that it takes $t/\delta t$ time steps to
reach the lattice position $r/\delta r$, given that the walker is at the
origin at time $t=0$. Whatever the microscopic dynamics, we assume that we
are given, or can work out, the set of probabilities $\{p_{j}(r)\}_{j=1}^{%
\infty }$ that the time between the first visit of the lattice site $%
r/\delta r$ and the first visit of the next position, $r/\delta r+1$, is $%
\mu_j \delta t$. Conceptually, these represent the probabilities of various
waiting times from the first visit of lattice site $r/\delta r$ until the
first visit to $r/\delta r+1$, i.e. the distribution of single-step waiting
times. It is then clear that the probability that it takes time $t$ for the 
walker to reach the lattice site $r+\delta r$ is equal to the probability that 
it takes time $t$ to reach lattice site $r$ and that the waiting time is zero,
plus the probability that it takes time $t-\delta t$ to reach site $r$ and that the 
waiting time is $\delta t$, plus the probability that the waiting time is 
$2\delta t$, . . .  so that the master equation is %
\begin{equation}
{f}(t/\delta t;r/\delta r+1)=\sum_{j=0}^{n}p_{j}(r)
{f}(t/\delta t-\mu_j;r/\delta r)\,,  
\label{first_visit}
\end{equation}%
or
\begin{equation}
{f}(t; r+\delta r)=\sum_{j=0}^{n}p_{j}(r)
{f}(t - \tau_j; r)\,.  
\label{a1}
\end{equation}%
This is the {\em first visit equation} \cite{boon} where
$p_j$ is the probability that it takes time $\tau_ j\,=\mu_j\,\delta t$ for
the particle to propagate from $r$ to $r + \delta r$, 
i.e. $\tau_j$ is the time delay between two successive first visits on the 
propagation strip for the path with probability~$p_j$. 
The sum is over all possible time delays, weighted by the probability~$p_j$,
and $n$ can be finite \cite{boon} or infinite \cite{buni_khlabys}, the two
formulations being equivalent, depending on whether the distribution of the 
delays is contained either in the  $\tau_ j$'s or in the $p_j$'s.

For one particular realization, the successive time delays are set by
a given spatial configuration of the time delayers, and the time
taken by the particle to perform a displacement from $r$ to $r + \delta r$ 
depends on that configuration. For an ensemble of realizations,
the distribution function of the time delays defines the average 
displacement time
\beqa
\la \tau \ra\,=\, \sum_{j=0}^n\,\,p_j\,\tau_j
\,=\, \sum_{j=0}^n\,\mu_j\,p_j\,\delta t
\,=\,\la\mu\ra\,\delta t \;,
\label{a2}
\eeqa
and the variance
\beqa
\la \tau^2\ra -\la \tau\ra ^2
&=&\left\{\sum_{j=0}^n \mu_j^2\,p_j\,-[\sum_{j=0}^n \mu_j\,p_j]^2\right\}\,
(\delta t)^2 \nm
&=&(\la \mu^2\ra -\la \mu\ra ^2)\,(\delta t)^2 \;,
\label{a3}
\eeqa
where $\mu_j=\tau_j/\delta t$ is the number of time steps during the 
time delay $\tau_j$. The general condition on the $p_j$ distribution 
is that its moments be finite.
For specific lattice dynamics (such as for the 1-D model described above)
$\mu_j$ is known analytically and the moments can be computed explicitly.
For instance, one can then show that the propagation velocity of Langton's
ant is $c = \delta r / \la \tau \ra = 2 \sqrt 2 / 104$ \cite{boon}.

\section{Propagation-Dispersion Equation}

The systems that we are discussing exhibit two time scales which correspond to
(i) a propagation process characterized by the average time necessary 
to complete a finite number of displacements $r/\delta r$ 
\beq
\label{a4}
{\rm E}[t_r]\,=\,\la\mu\ra\,r\,\frac{\delta t}{\delta r}\;,
\eeq
and (ii) the dispersion around this average value characterized by
the variance 
\beq
\label{a5}
{\rm Var}[t_r]\,=\,(\la \mu^2\ra \,-\,\la \mu\ra ^2)\,(\delta t)^2\,
\frac{r}{\delta r}\;.
\eeq
For finite $r$, these are finite quantities. Correspondingly we define
the following quantities that will be used in the hydrodynamic limit 
of Eq.(\ref{a1})
\beq
\label{a6}
\frac{1}{c}\,=\,
\la\mu\ra\,\frac{\delta t}{\delta r}\;,
\eeq
and
\beq
\label{a7}
\gamma\,=\,
(\la\mu^2\ra \,-\,\la\mu\ra ^2)\,\frac{(\delta t)^2}{\delta r}\;.
\eeq
$c$ ($\neq 0$) will be identified as the propagation speed  and 
$\gamma$ ($\geq 0$) will be identified as the dispersion coefficient.

The hydrodynamic limit, i.e. for $r / \delta r \gg 1$, can be obtained by
multi-scale expansion starting from Eq.(\ref{a1}) or by the generating function 
method (with application of the central limit theorem) 
\footnote{The reader is referred to  \cite{BGL1, BGL2} for the analytical 
computations.}. With the multi-scale expansion, one obtains the 
{\em propagation-dispersion equation} \cite{BGL1}

\beq
\label{a19}
\partial_r\,f(r,t)\,+\,\frac{1}{c}\,\partial_t f(r,t)\,=\,
\frac{1}{2}\,\gamma\,\partial_t^2 f(r,t)\;,
\eeq
and with the other method, one obtains its solution \cite{BGL2}
\beq
\label{a20}
f(r,t)\,=\,\sqrt{\frac{1}{2 \pi}}\,(\gamma\,r)^{-\frac{1}{2}}
\exp \,\left (-\,\frac{(t\,-\,\frac{r}{c})^2}{2\,\gamma\,r}\right )\;,
\eeq
with the the initial condition that at the origin, say at $r=0$,
$f(0,t)\,=\,\delta (t)$. Note that this condition is not restrictive
in that, if the initial distribution is given by some function
$f(t; r=0) = f_0(t)$, the solution is the result (\ref{a20}) convoluted
with $f_0(t)$. 

It is clear from \Eq{a19}, that $c$ is a propagation speed, and $\gamma$
is a transport coefficient expressing dispersion in time (instead of
space like in the classical Fokker-Planck equation for diffusion).
Equation \Eq{a19} is the propagation-dispersion equation governing the
first-passage distribution function of a propagating particle subject 
to time delays. Figure 2 illustrates these results.

\begin{figure}
\includegraphics[height=.5\textheight]{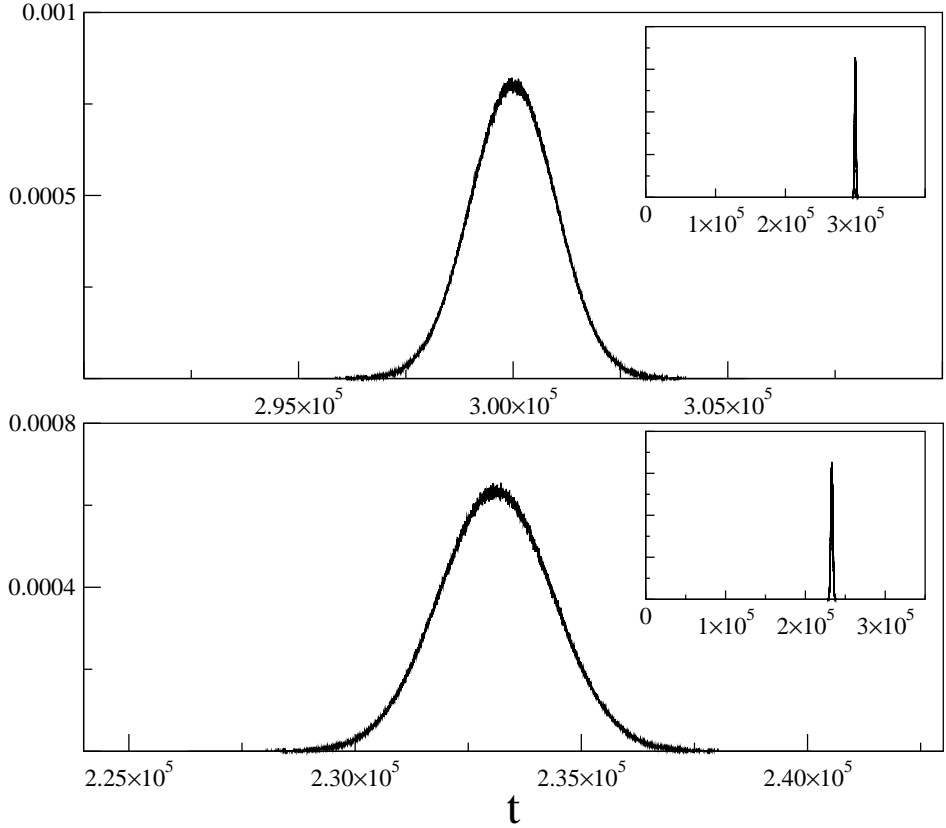}
\caption{Probability distribution $f(r=3 \times 10^4,\,t)$ 
based on general equation~\Eq{a1}. 
(a) time delays equally distributed for $j\,=0,\,1,\,\cdots , \,9$, 
with $p_j\,=\,0.1$; $c = 0.1$ and $\gamma = 33$; 
half-width $=\sqrt {2 \gamma r} \simeq 1.41\times 10^3$.
(b) time delays exponentially distributed: 
$p_j\,=\,C \,\exp{-\beta j}$, with $j=0,\,1,\,...,\,9$, $\beta\,=\,0.25$, 
and $C\,=\,[\sum_{j=0}^9 j]^{-1} = 1/45$; $c = 0.128 $ and $\gamma = 52.7$;
half-width $=\sqrt {2 \gamma r} \simeq 1.78 \times 10^3$.
The numerical simulation data and the analytical expression (Eq.(\ref{a20});
solid line, not visible) coincide perfectly. Insets show large scale representation.
Space and time are in automaton units.}
\end{figure}

Note that for the biased random walker in the continuum limit, the exact first 
passage time distribution is known \cite{feller, scher} to be
\begin{eqnarray}
f(t;r)=\frac{r}{\sqrt{2\pi D}\,t^{3/2}}
\exp \left( -\frac{\left( r-ct\right) ^{2}}{2Dt}\right)\;,
\label{14c}
\end{eqnarray}%
where $D$ is the spatial diffusion coefficient.
The difference between this expression and (\ref{a20}) is due to the fact 
that the latter is an approximation which is only valid for large $r$. 
In this regime, the exact result only gives a non-zero probability for 
$\left( r-ct\right) ^{2}/2Dt = {\cal O}(1)$ which implies 
$ct = r+ {\cal O}\left( \sqrt{2Dr/c}\right) 
=r\left( 1+ {\cal O}\left( \sqrt{2D/cr}\right) \right)$. 
So, for large $r$ we can use this approximation to write the exact 
distribution as
\begin{eqnarray}
f(t;r)=\frac{c^{3/2}}{\sqrt{2\pi D}\,r^{1/2}}
\exp \left( -\frac{\left( r-ct\right) ^{2}}{2Dr/c}\right) 
\left( 1+ {\cal O}\left( \sqrt{2D/cr}\right) \right)\;,
\label{14d}
\end{eqnarray}%
which, with $D/c^{3} = \gamma$, agrees with the large-distance 
result (\ref{a20}). We emphasize that Eq.(\ref{14c}) is exact in the continuum
limit, i.e. for vanishing $\delta r$ and $\delta t$, whereas the only restrictions on the general result (\ref{a20}) are that $r$ is large and that the first two moments of the elementary waiting time distribution, $\la \tau \ra$ and $\gamma$, exist. 
The latter condition precludes the limit of the symmetric random walker, 
$c\rightarrow 0$, for which  $\la \tau \ra = \delta r /c$ diverges (see (\ref{a2})
and (\ref{a6})).

\section{Temporal diffusion}

It follows from Eqs.\Eq{a5} and \Eq{a7}, that the dispersion coefficient
$\gamma$ is given by
\beq
\label{b11}
\gamma\,=\,\frac{\la t_r^2 \ra -\,\la t_r \ra^2}{r}\,
\eeq
which, for large $r$, is reminiscent of the classical expression for 
the diffusion coefficient: $D\,=\,\lim_{t\to \infty} \la r^2(t) \ra/2t$. 
Comparison of the two expressions shows interchange of space and time,
and measurements of the variance $\la t_r^2 \ra-\,\la t_r \ra^2$ should show
a linear dependence in terms of the distance 
with a slope equal to $\gamma$ in the same way
as the diffusion coefficient is obtained as the slope of the mean-square
displacement versus time in the long-time limit. 
 
An interesting case is the experimental study of the diffusion of a single 
particle in a 3-D random packing of spheres \cite{hulin} which describes
the motion of a particle through an idealized granular medium. 
Here one measures particulate transport and `dispersivity' which 
corresponds precisely to the quantity $\gamma$. The experimental data 
presented in \cite{hulin} show  that the mean square 
transit time of the particle through the medium is a linear 
function of the mean transit time (Figs.10 and 11 in \cite{hulin})  
itself a linear function of the percolating distance (Fig.2 in 
\cite{hulin}). This observation is a clear experimental illustration 
of the feature of Eq.(\ref{b11}). This experimental study also shows 
that the particle transit time is Gaussianly distributed in time (see Fig.9 in
\cite{hulin}) in accordance with the solution (\ref{a20}) of 
Eq.(\ref{a19}) (see Fig.2). 

A popular example where the concept of temporal diffusion is obviously
relevant is the Marathon. Each runner can be viewed as a particle
moving on a one-dimensional path - the race track - starting from a given
origin and heading towards the finish line, with time delays along 
its trajectory. Each such trajectory represents one realization of the 
dynamics, which generates a distribution approximated by a Gaussian
(as shown in Fig.3) whose first moment is the average time of arrival 
($\la \tau \ra \simeq 255$ min) with  
$c = 42.195  \times 60 /\la \tau \ra \sim 10$ (km/hour), and whose 
second moment  gives a measure of the dispersion coefficient 
$\gamma \simeq .055$ min$^2$/m.

\begin{figure}
\rotatebox{-90}{
\includegraphics[height=.5\textheight]{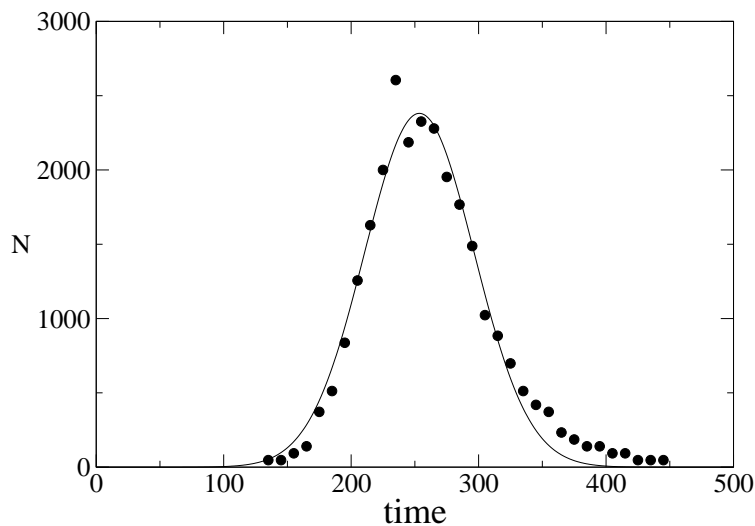}}
\caption{New York marathon (1996): distribution of arrival times
(in minutes; N =  number of runners). Data (black dots) and 
Gaussian fit (solid curve). The skewness indicates that all runners
are not subject to the same waiting time probability distribution.}
\end{figure}

The dispersion coefficient can be further expressed in terms of the
fluctuations in the {\em local propagation velocity} $v(r)$, a quantity with 
average value $c$. In fact it is the reciprocal local velocity which is
physically relevant: it is the time taken by the particle to 
propagate from position $r$ to $r + \delta r$ (divided by $\delta r$).
Then indeed
\beq
\label{b10}
\la t_r\ra \,=\,\la \int_0^r\,dr'\,\frac{1}{v(r')}\,\ra\,=
\,\int_0^r\,dr'\,\la\frac{1}{v(r')}\,\ra\,=\,\frac{r}{c}\,,
\eeq
which is consistent with the definition of the propagation speed.
It is then easy to compute the variance in terms of the reciprocal velocity 
fluctuations $\delta v^{-1}(r) = v^{-1}(r) - \la v^{-1}\ra =  v^{-1}(r) - c^{-1}$,
\beqa
\label{b14}
\la t_r^2\ra - \la t_r\ra^2 &=&\int_0^r\,dr'\,\int_0^r\,dr''
\la \delta v^{-1}(r') \, \delta v^{-1}(r'')\ra\,.
\eeqa
If the dynamics of the propagating particle is such that the 
correlation function on the r.h.s. of (\ref{b14}) is $\delta$-correlated, i.e. 
$\la \delta v^{-1}(r') \, \delta v^{-1}(r'')\ra\,=\,\phi_0\,
\delta (\frac{r'}{\xi}-\frac{r''}{\xi})$ with 
$\phi_0\,=\,\la (\delta v^{-1})^2\ra\,=\,\la \frac{1}{v^2}\ra -
\frac{1}{c^2}$, and where $\xi$ is the elementary correlation length,
it follows from \Eq{b11} and (\ref{b14}) that 
\beq
\label{c17} 
{\gamma}\,=\,\xi (\langle v^{-2} \rangle - c^{-2})\,,
\eeq 
that is ${\gamma}$ is the covariance of the reciprocal velocity 
fluctuations multiplied by the correlation length. This result is analogous 
to Taylor's formula of hydrodynamic dispersivity which is expressed as the 
product of the covariance of the velocity fluctuations with a 
characteristic correlation time \cite{taylor}.
Here $\gamma$ is the {\em temporal} dispersivity. 

In classical advection-diffusion phenomena, the control parameter is 
the P\'eclet number $P=UL/2D$, where $U$ denotes the mean advection 
speed, $L$, the characteristic macroscopic length, and $D$, 
the diffusion coefficient (see e.g. \cite{koplik}). The analogue for 
propagation-dispersion follows by casting Eq.\Eq{a19} in non-dimensional 
form 
\beq
\label{a13nd}
\partial_{\tt r}\,f(r,t)\,+\,\partial_{\tt t}\,f(r,t)\,=\,B^{-1}\,
\partial_{\tt t}^2\,f(r,t)\;;\;\;\;\; B=\frac{2T}{\gamma c}\;.
\eeq
Here ${\tt r}$ and ${\tt t}$ are the dimensionless space and time
variables: ${\tt r} = r(cT)^{-1}$ and ${\tt t} = t\,T^{-1}$, where
$T$ is a characteristic macroscopic time. $B$ is the control
parameter for propagation-dispersion: it is a measure of the relative
importance of propagation with respect to dispersion. 
Indeed, $B=\frac{2T}{\gamma c}= \frac{2 T^2}{\gamma}\frac{1}{cT}= L_D/L_P$,
i.e. the ratio of the characteristic dispersion length $L_D$ to the
characteristic propagation length $L_P$. At high values of $B$, i.e.
$L_D \gg L_P$, the distribution function is very narrow, and transport
over large distances ($r \geq cT$) is dominated by propagation.

\section{Generalized Propagation-Dispersion}

There are two important generalizations of the propagation-dispersion
equation. The first generalization is for temporal diffusive behavior in
inhomogeneous systems, i.e. for processes where the waiting time
probabilities depend on the location of the particle. The $p_j$'s are then
space dependent, and the propagation-dispersion equation 
becomes~\cite{BGL2} 
\begin{equation}
\frac{\partial }{\partial r}f(t,r)+\frac{1}{c(r)}\frac{%
\partial }{\partial t}f(t,r)=\frac{1}{2}\gamma \left(
r\right) \frac{\partial ^{2}}{\partial t^{2}}f(t,r)
\label{10}
\end{equation}%
with 
\begin{eqnarray}
\frac{1}{c(r)} &=&\frac{\partial }{\partial r}\tau(r) \,, 
\label{11a} 
\end{eqnarray}%
and
\begin{eqnarray}
\gamma \left( r\right)  &=&\frac{\partial }{\partial r}
\sigma^2 \left(r\right) \,,
\label{11b}  
\end{eqnarray}%
where
\begin{equation}
\tau \left( r\right)  =\delta t\sum_{k=1}^{r/\delta r}\sum_{j=0}^{\infty
}j p_{j}(\left( k-1\right) \delta r)\,,  
\label{3}
\end{equation}%
and 
\begin{eqnarray}
\sigma^2 (r) =  \left( \delta t\right) ^{2}
\sum_{k=1}^{r/\delta r}\left[
\sum_{j=0}^{\infty }p_{j}(\left( k-1\right) \delta r)j^{2}-\left(
\sum_{j=0}^{\infty }p_{j}(\left( k-1\right) \delta r)j \right) ^{2}\right]\,. 
\label{4}
\end{eqnarray}%
The solution of Eq.(\ref{10}) reads
\begin{equation}
f(t,r)=\int_{-\infty }^{\infty }\sqrt{\frac{1}{2\pi \sigma^2 (r)}}%
\exp \left( -\frac{\left( t-t' -\tau\left( r\right) \right) ^{2}}{%
2\sigma^2 (r)}\right) f_{0}\left( t' \right) dt' \,,  
\label{9}
\end{equation}%
where $f_{0}\left( t\right)\,=\, f(t; r=0)$. 
Buminovich and Khlabystova \cite{buni_khlabys} have studied 
models similar to the one-dimensional model described earlier in this
chapter, but in which the scatterers only change state after multiple scattering 
events \cite{bunimov}. In this case, the distribution of elementary waiting 
times becomes dependent on the lattice position, and the propagation speed 
and dispersion coefficient acquire a spatial dependence. Thus, while the 
distributions of first passage times are still Gaussian, they are not 
``diffusive'' in the usual sense since the inverse propagation speed and 
dispersion coefficient are not constants (Eqs.(\ref{10}-\ref{11b})).

The phenomena described  so far are for cases where the variance of the 
elementary time-delay processes exist. The second generalization is for the 
interesting class of similar, but more complex processes which are described 
by power-law distributions which do not possess second moments,
 e.g. $p_{j} \sim {\tau_j}^{-(1+\alpha)}$, in which case, for $0< \alpha \leq 1$, 
 the distribution appearing in the central limit theorem is no longer Gaussian 
(see e.g. \cite{gnedenko} and the appendix in \cite{berengut}). 

For the Pareto distribution
\begin{equation}
\label{pareto}
p(t)=\Theta \left( t-t_{0}\right) \frac{\alpha
t_{0}^{\alpha }}{t^{1+\alpha }} \;\;\; ; \;\;\; 0<\alpha <2,\;\alpha \neq 1\;,
\end{equation}
one can show that the propagation-dispersion equation becomes
\begin{equation}
\label{frac_eq}
\frac{\partial }{\partial r}f_{\alpha }\left( t;r/\delta r\right) =\left[ 
\frac{\alpha t_{0}}{\left( 1-\alpha \right) \delta r}\frac{\partial }{%
\partial t}-t_{0}^{\alpha }\frac{\Gamma \left( 1-\alpha \right) }{\delta r}%
\frac{\partial ^{\alpha }}{\partial t^{\alpha }}\right] f_{\alpha }\left(
t;r/\delta r\right)  \;,
\end{equation}%
where the fractional derivative can be defined through the Fourier
transformation
\beq
\label{frac_deriv}
\frac{\partial}{\partial t^{\alpha }}  f_{\alpha }(t) = 
\frac{1}{2 \pi} \int_{-\infty}^{+\infty} d\omega (-\imath \omega)^{\alpha}
\exp (-\imath \omega t) \tilde f (\omega) \;.
\eeq
Equation (\ref{frac_eq}) is the fractional propagation-dispersion equation.
For the special case $\alpha =1/2$, the equation becomes
\begin{equation}
\label{frac_eq_2}
\frac{\partial }{\partial r}f_{1/2}\left( t;r/\delta r\right) =\left[ \frac{%
t_{0}}{\delta r}\frac{\partial }{\partial t}-\frac{\sqrt{\pi t_{0}}}{\delta r%
}\frac{\partial ^{1/2}}{\partial t^{1/2}}\right] f_{1/2}\left( t;r/\delta
r\right)  \;,
\end{equation}
whose solution reads
\begin{equation}
\label{sol_eq2}
f_{1/2}\left( t;r/\delta r\right) =\frac{1}{2}\sqrt{t_{0}}\frac{\left( \frac{%
r}{\delta r}\right) }{\left( t+\frac{r}{\delta r}t_{0}\right) ^{3/2}}\exp
\left( -\frac{t_{0}\pi \left( \frac{r}{\delta r}\right) ^{2}}{4\left( t+%
\frac{r}{\delta r}t_{0}\right) }\right)  \;.
\end{equation}%
The fractional equation (\ref{frac_eq}) should be contrasted with 
the fractional Fokker-Planck equation which has been 
studied extensively for anomalous spatial diffusion \cite{klafter}.
The fractional propagation-dispersion equation (\ref{frac_eq}) is
new and is expected to be appropriate for the description and the 
analysis of non-Gaussian (anomalous) temporal diffusive processes.

\section{Comments}

There is an algebraic similarity in the structure of the 
propagation-dispersion equation (\ref{a19}) and that of the classical 
advection-diffusion equation~\cite{feller} which can be formally 
transformed into each other by interchanging space and time variables. 
It should be clear that the two equations describe different, but
complementary aspects of the dynamics of a moving particle.
Solving the propagation-dispersion equation answers the question of 
the time of arrival and of the time distribution around the average
arrival time in a propagation process. It is also legitimate to
ask the complementary question ``where should we expect to find the 
particle after some given time ?'' which should be long compared to 
the elementary time step, but short with respect to the average time 
of arrival. We will then observe spatial dispersion around some
average position which can be evaluated from the solution of the 
advection-diffusion equation. This observation stresses the 
complementarity of the two equations. 

Because the propagation-dispersion equation describes the space-time 
behavior of the {\em first passage} distribution function $f(r, t)$, i.e. 
the probability that a particle be for the first time at some position, 
it describes transport where a first passage mechanism plays an important 
role. So the equation should be applicable to the class of front-type 
propagation phenomena where any location ahead of the front will
necessarily be visited, the question being: {\em when} will a given point
be reached? 

Besides the examples discussed above, temporal diffusion is also 
encountered in shock propagation in homogeneous or inhomogeneous 
media \cite{schock} or packet transport in the Internet \cite{internet}. 
As the propagation-dispersion equation 
is for the first-passage time distribution, it should also be suited for 
the description of transport driven by an input current in a disordered 
random medium \cite{kehr}. In the area of traffic flow, there are typical 
situations where cars moving on a highway from location A to location 
B, are subject to time delays along the way, and -- with the assumption 
that all cars arrive at destination --  one wants to evaluate 
the time of arrival \cite{traffic}. Financial series as
in the time evolution of stock values are another example
\cite{financial}: over long periods of time (typically years) one
observes a definite trend of increase of, for instance, the  value
of the dollar. So any preset reachable value will necessarily be attained, 
the questions being: when? and what is the time distribution around the 
average time for the preset value? While the classical question is: 
after such or such period of time, which value can one expect?, there
might be instances where the reciprocal question should be considered.  
Because of the generality of the propagation-dispersion equation, it
should be expected that, either in its simple form \Eq{a19} or in its 
generalized forms, \Eq{10} and \Eq{frac_eq}, the equation be applicable to 
a large class of first-passage type problems in physics and related domains.


\begin{theacknowledgments}
This chapter is based on a series of articles co-authored with 
P. Grosfils, J.F. Lutsko, E.G.D. Cohen, and L.A. Bunimovich.
It is my pleasure to acknowledge their stimulating and fruitful 
collaboration.
\end{theacknowledgments}





\IfFileExists{\jobname.bbl}{}
 {\typeout{}
  \typeout{******************************************}
  \typeout{** Please run "bibtex \jobname" to optain}
  \typeout{** the bibliography and then re-run LaTeX}
  \typeout{** twice to fix the references!}
  \typeout{******************************************}
  \typeout{}
 }

\end{document}